# Comparative Analysis of Human Movement Prediction: Space Syntax and Inverse Reinforcement Learning

**Soma Suzuki**

Center for Advanced Spatial Analysis
University College London

## Abstract

Space syntax matrix has been the main approach for human movement prediction in the urban environment. An alternative, relatively new methodology is an agent-based pedestrian model constructed using machine learning techniques. Even though both approaches have been studied intensively, the quantitative comparison between them has not been conducted. In this paper, comparative analysis of space syntax metrics and maximum entropy inverse reinforcement learning (MEIRL) is performed. The experimental result on trajectory data of artificially generated pedestrian agents shows that MEIRL outperforms space syntax matrix. The possibilities for combining two methods are drawn out as conclusions, and the relative challenges with the data collection are highlighted.

## I. Introduction

The problem of predicting human movement in urban environments features prominently in urban planning as the acquired model is considered as the artificial laboratory that enables us to test ideas and hypothesis that are not easy to explore in the real world (Torrens 2015). Historical approach to infer pedestrian movement is a methodology developed in space syntax community, in which city is represented as a connectivity graph so as to explain human movement rate in each space (Hillier 1993). Even though it turned out that some defined metrics are correlated to human movement rates in the city, it is most unlikely that humans move merely based on the number of links each space has. An alternative promising approach to predict pedestrian movement is to generate human-like agent through learning humans' preferences on road choice given observational trajectory data. This approach is known as Inverse Reinforcement Learning (IRL) in Markov Decision Process (MDP) in machine learning community (Ng and Russell 2000). IRL has been already studied intensively for predicting human movement (Ziebart 2008). However, the valid comparison between IRL and space syntax matrix has not been performed.

This paper conducts a comparative analysis of two methods for predicting human movement in the urban environment. In section II, space syntax approach for modeling human movement is explained. Next, section III introduces Maximum Entropy Inverse Reinforcement Learning (MEIRL) to generate human-like agents. Then, the quantitative comparison between space syntax matrix and MEIRL is conducted in section IV. Finally, the possibility of combining space syntax matrix and MEIRL and the future work is discussed as conclusion in section V.

## II. Space Syntax Metrics

First developed by Hillier in University College London (Hillier 1984), space syntax has grown into an independent research area. Primarily, space syntax is a methodology for investigating spatial complexities in an attempt to identify its particular structure that resides at the level of the entire configuration. The methodology is based on the theory that the form-function relation in cities reflects the structural properties of its configuration (Hillier 1998). Through enormous empirical study, it turns out that space syntax metrics can capture human movement (Penn 1998) and has been widely adopted on human movement prediction. In space syntax approach, the city is first represented as connectivity graph, consisting of nodes representing spaces and edges if the corresponding spaces are intersected as shown in Figure 1. Then, individual spaces are ranked based on space syntax matrix for inferring human movement rates.

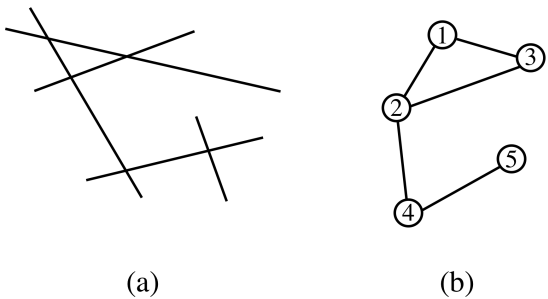

Figure 1:
A fictive city (a) and connectivity graph (b)

Among various space syntax metrics, researchers in space syntax community claim that pedestrian movement is most well-described by local integration. As illustrated below, local integration is a normalized closeness centrality initially developed in social network analysis (Scott 2000).

*Space Syntax Metrics: Local Integration*

Local integration is based on the concept of depth, which assesses how far it is from a node to other nodes within 2 steps. For any particular node in the connectivity graph, the distance (or step) $d_{ij}$ between point and point j denotes the length of the shortest path linking them, if any; otherwise $d_{ij} = \infty$. For all points, the following set of axioms holds:

- $d_{ij} \geq 0$, with $d_{ij} = 0$ only if $i = j$
- $d_{ij} = d_{ji}$
- $d_{ij} + d_{jk} \geq d_{ik}$

To obtain local integration, closeness centrality of a node needs to be calculated first. Closeness centrality measures the centrality of a node in a network, which is obtained as the reciprocal of the sum of the number of the steps to other nodes. Thus closeness centrality of node *i* within two steps is calculated as follows:

$$C_i = \frac{1}{\sum_j^k d_{ij}} \quad (if\ d_{ij} \leq 2)$$

where *k* represents the number of nodes in the connectivity graph. For instance, node 1 has two nodes in step one and 1 node in step two, thus the sum $1 / (2 \times 1 + 1 \times 2)$ indicates the closeness centrality of node 1 considering two steps. Then, the reciprocal of closeness centrality divided by the number of nodes involved minus one leads to local mean depth *MD*. Finally, by standardizing the variation of mean depth *MD* between zero and one, relative asymmetry *RA* is obtained as the local integration of a node (Hillier 1984).

$$MD_i = \frac{C_i}{k-1}$$

$$RA_i = \frac{2(MD_i - 1)}{k-2}$$

Thereafter individual space is ranked according to the relative asymmetry *RA* to predict pedestrian movement rate in each space. Even though why the closeness centrality or local integration can be adopted to infer human movement is not clearly justified, a vast amount of empirical data proves that local integration works very well for predicting pedestrian movement in the city.

## III. Maximum Entropy IRL

The alternative approach to predict pedestrian movement is to learn human preference and generate human-like agent. In this section, Maximum Entropy Inverse Reinforcement Learning (MEIRL) is introduced as the methodology for learning pedestrians' preferences. Inverse Reinforcement Learning (IRL) is a problem in Markov Decision Process (MDP), that is, the problem of extracting a reward function given observed, optimal behavior (Ng and Russell 2000).

*Markov Decision Process*

A finite MDP is a tuple ($S$, $A$, $P_{sa}$, $\gamma$, $R$), where

- $S$ is a finite set of $N$ states $\{s_1, s_2, ..., s_n\}$.
- $A$ is a set of $K$ actions $\{a_1, a_2, ..., a_k\}$.
- $P_{sa}$ is state transition probabilities upon taking action $a$ in state $s$.
- $\gamma \in [0,1]$ is the discount factor.
- $R : \mathbb{S} \times \mathbb{A} \to \mathbb{R}$ is the reward function that depends on state and action.

The classical problem of MDP is to find the optimal policy $\pi^* : \mathbb{S} \to \mathbb{A}$ such that expected reward is maximized.

*Inverse Reinforcement Learning*

Conversely, Inverse Reinforcement Learning (IRL) problem consists of finding the reward function from an observed policy. More specifically, given a finite space $S = \{s_1, s_2, ..., s_n\}$, set of actions $A = \{a_1, a_2, ..., a_k\}$, transition probabilities $P_{sa}$, a discount factor $\gamma \in [0,1]$ and a policy $\pi : \mathbb{S} \to \mathbb{A}$, the goal of IRL is to find a reward function such that $\pi$ is an optimal policy $\pi^*$. The acquired reward function explains the intrinsic preference of policy demonstrator, or expert agents. Therefore, learning reward function through IRL enables us to generate artificial agents that behave in such a way that pedestrians "would" do in the city. Training data for IRL is a set $D = \{\zeta_1, \zeta_2, ..., \zeta_m\}$ of $m$ independent trajectories sampled from the expert's (pedestrians) policy $\pi_E$.

Learning optimal policy from experts' demonstrations is called imitation learning. In the imitation learning setting, the targeted reward function is approximated based on path feature counts, $F_\zeta = \sum_{s_j \in \zeta} F_{s_j}$, which is the sum of the state features along the path. Thus, given the reward weight $\theta$ that linearly maps each state feature $F_{s_j}$ to a state reward value, the total reward value of a trajectory is:

$$R_\theta(F_\zeta) = \theta^T F_\zeta = \sum_{s_j \in \zeta} \theta^T F_{s_j}$$

The problem of imitation learning, therefore, consists of approximating the reward weight $\theta$ by matching feature counts between expert agent's policy and the learner agent's policy.

*Maximum Entropy IRL*

To obtain the distribution of path that performs one-to-one correspondence to feature counts, the principle of maximum entropy (Jaynes 1957) is employed. In other words, the adopted distribution of path does not exhibit any additional preferences for some paths beyond feature counts (Ziebart 2008). The probability of a path $\zeta_i$ to be chosen by an agent under reward weight $\theta$ is given as follows:

$$P(\zeta_i|\theta) = \frac{1}{Z(\theta)} exp(\theta^T F_{\zeta_i})$$

$$Z(\theta) = \sum_{\zeta \in D}^{m} exp(\theta^T F_\zeta)$$

where the partition function $Z(\theta)$ is a normalized constant. Estimating optimal reward weight $\theta *$ implies maximizing the likelihood of the observed data $D$ under the maximum entropy probability distribution illustrated above. The optimal reward weight $\theta *$ is then obtained using the gradient of the (log) likelihood.

$$\theta* = arg \max_\theta L(\theta)$$

$$L(\theta) = \frac{1}{M} \sum_{\zeta \in D}^{m} log P(\zeta|\theta)$$

$$= \frac{1}{M} \sum_{\zeta \in D}^{m} log \frac{1}{Z(\theta)} exp(\theta^T F_\zeta)$$

$$= \frac{1}{M} \sum_{\zeta \in D}^{m} \theta^T F_\zeta - log Z(\theta)$$

$$= \frac{1}{M} \sum_{\zeta \in D}^{m} \theta^T F_\zeta - log \sum_{\zeta \in D}^{m} exp(\theta^T F_\zeta)$$

Thus, the gradient of the log-likelihood is

$$\nabla_\theta L(\theta) = \frac{1}{M} \sum_{\zeta \in D}^{m} F_\zeta$$

$$- \frac{\sum_{\zeta \in D}^{m} exp(\theta^T F_\zeta)}{\sum_{\zeta \in D}^{m} exp(\theta^T F_\zeta)} \frac{d\theta^T F_\zeta}{d\theta}$$

$$= \frac{1}{M} \sum_{\zeta \in D}^{m} F_\zeta - \sum_{\zeta \in D}^{m} P(\zeta|\theta) F_\zeta$$

$$= \widetilde{F} - \sum_{s \in \zeta} P(s|\theta) F_s$$

where the $\widetilde{F}$ is the expected state feature count of an expert agent. Therefore, the gradient is obtained as the difference between the expected state feature counts of expert agents and the expected state feature counts of learner agents, which can be expressed in terms of expected state visitation frequencies $P(s|\theta)$. A pseudocode is presented below.

---

**Pseudocode** Maximum Entropy IRL

---

$\theta \leftarrow prng()\ and\ gather\ D$

**for** $n = 1 : N$ **do**

  $R_\theta \leftarrow RewardUpdate(\theta)$

  $\pi_\theta \leftarrow PolicyUpdate(R_\theta)$

  $P(s|\theta) \leftarrow FrequencyCompute(\pi_\theta)$

  $\nabla_\theta L(\theta) \leftarrow GradCompute(P(s|\theta))$

  $\theta \leftarrow ParamUpdate(\nabla_\theta L(\theta))$

---

With the optimal reward function parameter $\theta^*$, generated agents are assumed to possess the preference similar to real pedestrians. Therefore, the state visitation frequencies $P(s|\theta^*)$ under the optimal reward function parameter $\theta^*$ is used to rank individual spaces for human movement prediction.

## IV. Experiment

In this section, the comparison between space syntax matrix and MEIRL is conducted given the trajectory dataset generated by using an open source vendor Open Street Map (OSM).

*Dataset*

To acquire plausible model, the vast amount of pedestrian trajectory, or training data is essential. In this experiment, the artificially generated pedestrians are adopted as collecting dataset pertaining to individual walking trajectory is very challenging (Torrens 2011). While the current pervasiveness of mobile devices enables the collection of mobile location data at the unprecedented scale and granularity, the challenge invariably remains in associating movement path with the reason "why" people moved as the recorded trajectory. To illustrate, the GPS data of those who are commuting to her office and those who are on the way to a nightclub cannot be used together as the training data to model pedestrian movement: the former will most likely choose the shortest path and the latter will likely be a random walk. With the purpose of facilitating machine learning approach toward human movement prediction in the city, the API is developed so as to be a pipeline for generating ready-to-be-used training dataset. Assuming that people walking along the shortest route are moving in order merely to reach the goal point, this API generates the shortest path from an arbitrary start point to a goal point in the selected area. The brief process flow of API is shown in Figure 2.

First, geographic data on OSM is downloaded in the selected area by using R package Osmar as illustrated in Figure 3. Then, as only the street data is needed for the trace generation, the street network is extracted from the downloaded geographical data as shown in Figure 4. It is done by selecting data that is tagged as "highway" which is the main key used for identifying any kind of road, street or path on OpenStreetMap. Next, the highway-osmar object is converted into a graph object in order to compute the shortest path. In the graph representation, each node represents an intersection and edge corresponds to a road as shown in Figure 5. Finally, using R package Igraph, the shortest path between randomly specified starting nodes and goal nodes are computed based on Dijkstra's shortest path algorithm. 400 trajectories dataset are generated with the random start and goal point as illustrated in Figure 6: 300 training set and 100 testing set. The constructed connectivity graph is also served to define Markov Decision Process for MEIRL methodology.

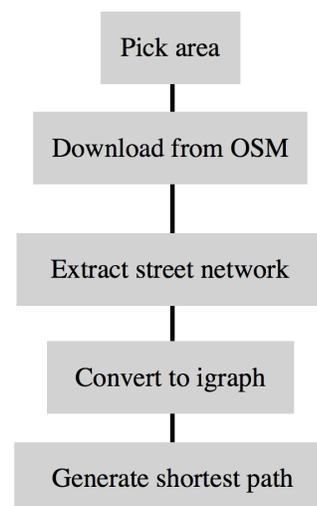

Figure 2: Process of built API

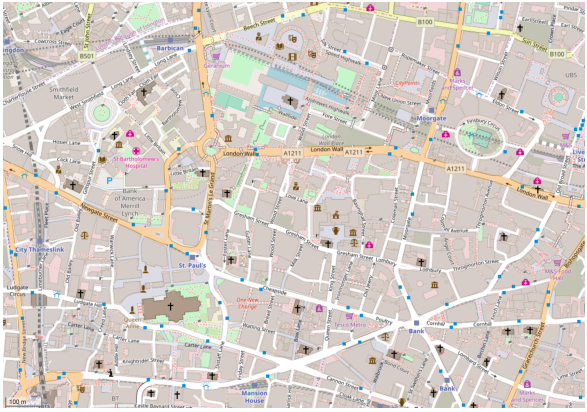

Figure 3: Selecting area in Open Street Map

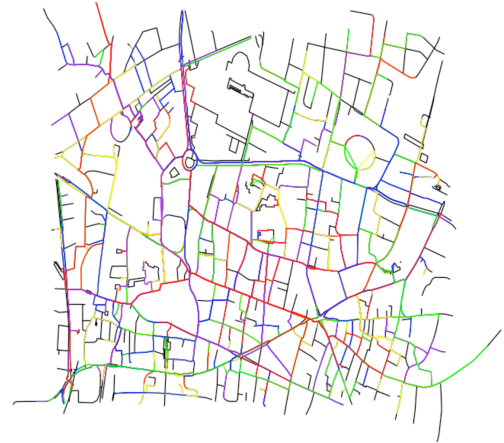

Figure 6: Generated training data. Different colors are introduced only for distinguishing each path.

Finally, MEIRL is conducted. The evolution of feature counts difference during the training is presented in Figure 7. In the experiment, the learning rate is set to a large value and gradually reduce as optimization progress. MEIRL performed well, showing quick convergence. Thereafter, the correlation between space syntax matrix and the test data and the correlation between the state visitation frequencies by MEIRL and the test data are compared.

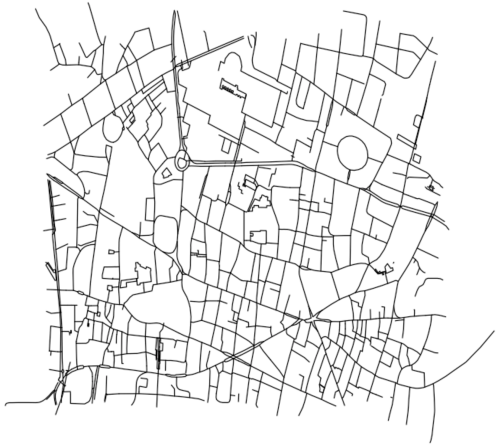

Figure 4: Extracted street network

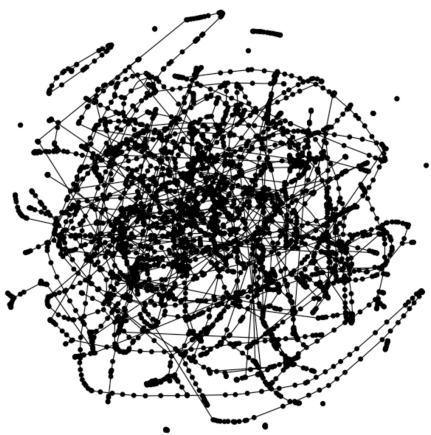

Figure 5: Connectivity graph

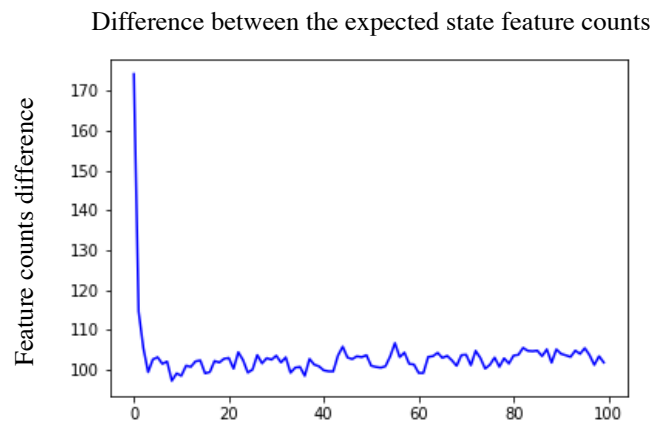

Figure 7: Evolution of feature counts difference

As a result, MEIRL showed stronger correlation 0.49 than 0.45 of space syntax matrix, meaning state visitation frequencies by MEIRL captures human movement in the city better than space syntax matrix.

## V. Conclusion

The MEIRL approach successfully performed more accurate prediction than space syntax matrix. However, I believe space syntax will be invariably useful in human movement prediction, employed together with presented agent-based approach. A major challenge in applying reinforcement learning in multi-agent model is how to manage the explosive computational cost as the state-action space grows exponentially with the number of agents and the learning becomes prohibitively slow. To alleviate this drawback, as known as structural reinforcement learning it is possible to reduce the size of the state-action space by supplying the model with the partial, but fundamental pedestrian movement in the form of space syntax graph. Future work may involve employing actual raw GPS data with the development of pipeline where the pedestrian trace data is plausibly associated with the label that explains the motivation of the corresponding movement.